# The aging effect in evolving scientific citation networks

Feng Hu[1,2,3] · Lin Ma[4] · Xiu-Xiu Zhan[4,5] · Yinzuo Zhou[4] · Chuang Liu[4] · Haixing Zhao[1,2,3] · Zi-Ke Zhang[4,6]

**Abstract**
The study of citation networks is of interest to the scientific community. However, the underlying mechanism driving individual citation behavior remains imperfectly understood, despite the recent proliferation of quantitative research methods. Traditional network models normally use graph theory to consider articles as nodes and citations as pairwise relationships between them. In this paper, we propose an alternative evolutionary model based on hypergraph theory in which one hyperedge can have an arbitrary number of nodes, combined with an aging effect to reflect the temporal dynamics of scientific citation behavior. Both theoretical approximate solution and simulation analysis of the model are developed and validated using two benchmark datasets from different disciplines, i.e. publications of the American Physical Society (*APS*) and the Digital Bibliography & Library Project (*DBLP*). Further analysis indicates that the attraction of early publications will decay exponentially. Moreover, the experimental results show that the aging effect indeed has a significant influence on the description of collective citation patterns. Shedding light on the complex dynamics driving these mechanisms facilitates the understanding of the laws governing scientific evolution and the quantitative evaluation of scientific outputs.

✉ Xiu-Xiu Zhan
  zhanxxiu@gmail.com

✉ Yinzuo Zhou
  zhouyinzuo@163.com

✉ Zi-Ke Zhang
  zkz@zju.edu.cn

1 School of Computer, Qinghai Normal University, Xining 810016, China

2 Key Laboratory of Tibetan information processing, Ministry of Education, Xining 810008, China

3 Tibetan information processing and Machine Translation Key Laboratory of Qinghai Province, Xining 810008, China

4 Alibaba Research Center for Complexity Sciences, Hangzhou Normal University, Hangzhou 311121, China

5 Faculty of Electrical Engineering, Mathematics and Computer Science, Delft University of Technology, 2628 CD Delft, The Netherlands

6 College of Media and International Culture, Zhejiang University, Hangzhou 310058, China

## Introduction

Scientific citation networks have provided a versatile and efficient tool to understand the structure and evolution of scientific progress (Martin et al. 2013; Wang et al. 2013; Leydesdorff 1998; Cronin 1984; Liu et al. 2018; Shi et al. 2019) by depicting topological interactions between academic publications and the propagation of scientific memes (Kuhn et al. 2014; Strogatz 2001; Shen et al. 2014), facilitating the emergence of a new research paradigm, the science of science (Fortunato et al. 2018; Zeng et al. 2017; Niu et al. 2016). The number of citations of a publication is a significant source to quantify its importance (Wei et al. 2013; Wang et al. 2013; Hirsch 2005). A vast class of evolving theoretical models has appeared to facilitate the understanding of the citation network. The first mathematical attempt was Price's model (Price 1965), which depicts papers as nodes and citations as links. Price also proposed a cumulative advantage process (Price 1976) to illustrate the *rich-get-richer* phenomenon (Adamic and Huberman 2000) (papers with more citations are more likely to be cited in the future) in citation networks, which is also known as the preferential attachment mechanism (Barabsi and Albert 1999), i.e., the *BA* model, in the field of complex networks, resulting in a power-law degree distribution. Despite its great success in addressing the underlying dynamics of network evolution, many empirical studies have shown that the power-law exponents of real-world data are much smaller than in the original *BA* model (Newman 2001; Jeong et al. 2003; Zhao et al. 2013), or they cannot be well fitted by a simple scaling function (Ucar et al. 2014; Lehmann et al. 2003; Shibata et al. 2008), suggesting that pure preferential attachment is not enough to describe the citation process. One missing factor is the aging effect, which considers that article influence is well correlated with publication time. On the one hand, one would expect early publications in a particular field to be highly cited, showing a strong *first-mover* effect (Newman 2009). On the other hand, studies are more likely to cite newly published papers to survey recent advances (Wei et al. 2013; Redner 1998; Leicht et al. 2007), exerting a strong impact on the shape of the citation distribution (Newman 2014; Dorogovtsev and Mendes 2000). Therefore, the temporal effect plays a significant role in modeling growing citation networks (Dorogovtsev and Mendes 2002; Medo et al. 2011).

Although *aging* is widely accepted as a key element influencing citation patterns, details of its involvement in the citation process remain unknown. Classical studies usually adopt graph theory to represent articles as nodes and citations as either directed or undirected links. This definition usually treats citations as pairwise relations. Comparatively, an alternative method, *hypergraph theory* (Berge 1973, 1989), which allows a so-called *hyperedge* to connect an arbitrary number of vertices, as opposed to two in regular graphs, has been proposed to solve the issue (Bashan et al. 2012; Johnson 2006; Menichetti et al. 2014). Due to its universal properties, the hypergraph has been widely applied to a vast class of structures, including social networks (Seidman 1981; Estrada and Rodrĺguez-Velzquez 2006; Chan and Hsu 2010), reaction and metabolic networks (Temkin et al. 1996; Krishnamurthy et al. 2003; Klamt and Gilles 2004), protein networks (Sarkar and Sivarajan 1998; Ramadan et al. 2004), food webs (Sonntag and Teichert 2004), social tagging networks (Ghoshal et al. 2009; Zlatić et al. 2009; Zhang and Liu 2010; Wang et al. 2010), scientific collaboration networks (Chakraborty and Chakraborty 2013; Hu et al. 2013), data and knowledge mining (Ozdal and Aykanat 2004; Liu et al. 2014; Guo and Zhu 2014), and more (Gallo et al. 1993; Konstantinova and Skorobogatov 2001; Bretto et al. 2002; Carstens 2014).

To more clearly illustrate the citation process, we present a dynamic hypergraph model that considers the aging effect. We compare the model results to two real citation datasets, including the American Physical Society (*APS*) and Digital Bibliography & Library Project (*DBLP*), and find good agreements. Further empirical analysis shows that the collective attraction of articles will decay in an exponential form.

## Materials and methods

In this section, we shall describe datasets and propose the evolving model. Traditional models of citation networks are usually described based on regular graphs, where one new node connects the old ones according to their respective connectivity (Bianconi and Barabsi 2001a, b). However, such a definition has some limitations in illustrating citation networks, (i) it only considers the pairwise relationship between the publication and any one of its references, neglecting to take into account the publication and all its the references as a whole; (ii) some highly-cited publications earn more citations according to the *rich-get-richer* effect (Adamic and Huberman 2000), and some new yet interesting papers may also be attractive, especially with the fast reading of the internet era (Ucar et al. 2014). We propose an evolving model with a hypergraph structure to address those limitations.

### Hypergraph evolutionary model

A hypergraph $H$ can be depicted by $H = (V, E)$, where $V = \{v_1, v_2, ..., v_N\}$ is a set of nodes, and $E = \{E_1, E_2, ..., E_e\}$ is a set of hyperedges consisting of arbitrary number of nodes. Each hyperedge represents the published paper, and nodes represent its references. $E_i \neq \phi$ and $\bigcup_{i=1}^{e} E_i = V$. Fig. 1 shows a simple example of a hypergraph. Analogous to regular networks, the hyperdegree is defined as the number of hyperedges connecting to the corresponding node. In our model, we use the preference attachment with aging effect to characterize the citation behavior of groups of nodes (papers). The aging effect is universal in the network evolution process. For example, the time span for both scientists and actors is finite. Thus, researchers have studied how different types of aging would affect the generated network structures (Zhu et al. 2003; Hajra and Sen 2006). Among the types of aging effect, the commonly used factors are exponential and power-law decay factor. We choose to use the representative power-law decay factor, $\tau^{-\alpha}$, where $\tau$ is the node age. We illustrate the hypergraph network generation process as follows:

– Initially, there is only one hyperedge, including one paper and all its $M_0$ references in the system (Fig. 2a);

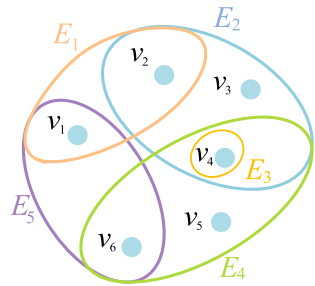

**Fig. 1** (Color online) Illustration of a typical hypergraph with six nodes and five hyperedges, including hyperedge $E_1$ with $\{v_1, v_2\}$, $E_2$ with $\{v_2, v_3, v_4\}$, $E_3$ with $\{v_4\}$, $E_4$ with $\{v_4, v_5, v_5\}$, and $E_5$ with $\{v_1, v_6\}$.

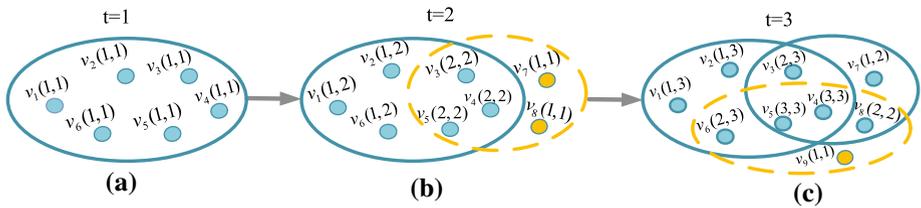

**Fig. 2** (Color online) Illustration of the evolutionary process of the model. Ellipses and circles respectively represent the hyperedges and nodes. $v_i(k, \tau)$ indicates that the hyperdegree of node $v_i$ with age $\tau$ is $k$. The orange dashed line and circles respectively represent newly added hyperedges and nodes at the corresponding time step

- At each time step, one new paper is published with $L$ references, among which are $m$ "existing nodes" ($m \leqslant L$) chosen from the system at time $t$ with probability

$$\Pi(i,t) = \frac{k_i(t)\tau_i^{-\alpha}}{\sum_j k_j(t)\tau_j^{-\alpha}}, \quad (1)$$

where $\tau_i = t - t_i + 1$ is the age of node $i$, $t_i$ is the time when node $i$ is introduced, $k_i(t)$ is the hyperdegree of node $i$ at time $t$, and $\alpha$ is a tunable parameter that indicates the strength of the aging effect (see Fig. 2b, c);
- The rest of the $L - m$ nodes are considered "new nodes" that have not been cited before;
- The above steps are repeated until the system achieves a considerable scale.

## Data description

We test the model with two real-world datasets. One is a collection of all papers published by the American Physical Society (*APS*). A total of 463,442 papers were published from 1893 to 2009, with 4,708,753 citations (details on the dataset can be accessed at https://publish.aps.org/datasets). The other dataset is a collection of all papers published by the Digital Bibliography & Library Project (*DBLP*), which focuses on computer science. There were 564,705 papers published from 1954 to 2013, with 4,191,677 citations (details on the dataset can be accessed at https://aminer.org/citation).

## Results and discussion

In this paper, we attempt to construct a hypergraph structure citation network model by introducing the aging effect. Based on this model, we show several results by observing different values of the parameter $\alpha$. We validate the model using empirical data.

### Model analysis

Eq. (1) is usually adopted to describe the temporal effect on simple graphs (Dorogovtsev and Mendes 2000; Medo et al. 2011), while the full hypergraph structure is rarely considered. We first investigate the impact of different aging values $\alpha$ on the hyperdegree

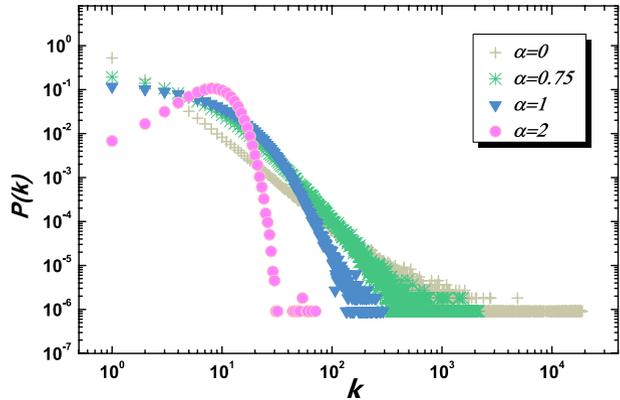

**Fig. 3** (Color online) Simulation results of hyperdegree distribution for different values of $\alpha$

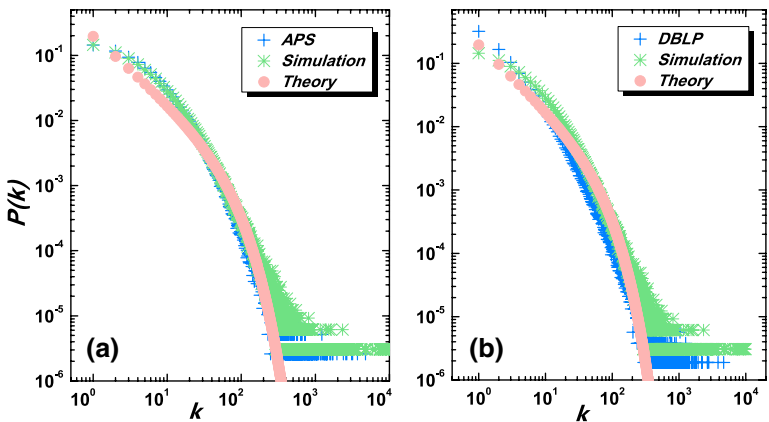

**Fig. 4** (Color online) Hyperdegree distributions of: **a** *APS*, and **b** *DBLP*, in which *APS* (*DBLP*) data (blue crosses), simulation (green stars) and theory analysis

distribution, given by Eq. (12) (see Appendix for analytical details). To provide a comparative study, we simulate four different metrics to characterize the properties of the citation distribution. Apart from the parameter $\alpha$, we set $L = 10$ and $m = \{6, 7, 8, 9, 10\}$, with the result that $q_m = \{0.05, 0.05, 0.05, 0.2, 0.65\}$. The expected value of the number of "old nodes" is $M = \sum_{m=1}^{L} mq_m = 9.35$, indicating that the expected value of "new nodes" is 0.65. Here, we consider the region $\alpha \geq 0$, since only this region seems to be of real significance. Fig. 3 clearly shows that the shape of the hyperdegree distribution is affected by the aging factor $\alpha$. For $\alpha = 0$, no aging effect is taken into account, and the present model degenerates to the classical BA model (Barabsi and Albert 1999), where a straight power-law distribution will be obtained. For $\alpha > 0$, the attraction of large-degree papers is suppressed as $\alpha$ increases, and a moderate hyperdegree distribution emerges. In the extreme case, $\alpha \to +\infty$, the hyperdegree distribution will follow a Poisson distribution according to Eq. (1), as only the most published papers will be cited.

In addition, we apply the model to two real citation networks, the *APS* and *DBLP* datasets (data details are described in Materials and Methods), to further evaluate the aging effect. Fig. 4 shows the consistency among the data, simulation, and theoretical results. In order to obtain the optimal $\alpha^*$ to fit the data, we use Kolmogorov-Smirinov (KS) test (Press et al. 1992; Clauset et al. 2009) to statistically validate the differences between the cumulative distributions of the real-data and simulation over all possible parameter space. In a KS statistic test, the goodness-of-fit value is defined as $D = max|f(k) - g(k)|$, where $f(k)$ and $g(k)$ are respectively the two cumulative distributions. Then the *p*-value is defined as the fraction of the synthetic distances that are larger than the empirical distance. Generally, if *p*-value is large enough (e.g. close to 1), then the difference between the two distributions can be attributed to statistical fluctuations. Instead, the two distributions will not be regarded identical if *p*-value is too small (e.g. close to 0), hence the proposed model will not be a plausible one to the corresponding data. With the help of KS test, we obtain optimal values of *α* for modeling *APS* and *DBLP* are $\alpha^*_{APS} = 0.75$ and $\alpha^*_{DBLP} = 0.5$, respectively. The goodness-of-fit values are $D_{APS} = 0.077$ and $D_{DBLP} = 0.053$, and *p*-values are $p_{APS} = 0.745$ and $p_{DBLP} = 0.976$ respectively for *APS* and *DBLP*, indicating that it can be accepted that the hyperdegree distributions of the data and model come from the same distribution, hence the proposed model can well explain both real datasets under respective optimal parameter *α*. Consequently, we use these values in the subsequent analysis. Fig. 4a, b respectively show the results from *APS* and *DBLP*. Note that the simulation results from the *APS* and *DBLP* data are obtained with $\alpha^*_{APS} = 0.75$ and $\alpha^*_{DBLP} = 0.5$, respectively, which are both greater than 0, indicating that preferential attachment is not the unique mechanism in facilitating the complex citation pattern. In addition, the positive *α* suggests that recent publications still have a great chance to be attractive to academicians.

To further understand how aging affects citation patterns, we empirically compare the present model to the observed data from two perspectives. First, we focus on the hyperdegree distribution of different *ages* along with the dynamic evolution process. For comparison, we divide both the simulated and empirical data into six groups, each ignoring previous publications and citations of the examined time. For example, in Fig. 5a, for articles published from the year 2000 (pink stars), we only count publications and corresponding citations from the year 2000 and eliminate all references before that year. In Fig. 5, both empirical and modeled data show that the shorter the observed period the more heterogeneous the distribution. Comparing the effects of the aging factor $\alpha > 0$ of the proposed model (see Fig. 5c, d), the results suggest that early publications benefit much more from a positive *α*, hence exerting more long-term influence. For short-term observations, although authoritative papers may dominate a field, new findings still can draw attention from the scientific world, as the fitted $\alpha = 0.75$ and $0.5$ are not infinitely large.

## Collective attraction

Subsequently, we turn to possess a comprehensive understanding of the notable phenomenon that all of the aforementioned results can be well fitted with empirical data with the designed decay factor parameter. The presented *α* indicates the decay rate of papers' attraction, especially the influence of hot papers, where a larger *α* suggests a quicker decay. To observe the decay speed of real data, we adopt a metric of collective attraction (Wu and Huberman 2007; Zhang et al. 2008),

$$r = \frac{E(logN_t) - E(logN_{t-1})}{E(logN_1) - E(logN_0)}, \qquad (2)$$

where $N_t$ is the number of citations of examined papers at time *t*, and $E(\cdot)$ is the expected value of $(\cdot)$. We set the year 1970 as $t_0$ for empirical data. Fig. 6 shows that the attraction of early publications from empirical data of two representative disciplines (physics and computer science) would decay exponentially with the form $r = Y_0 + A_0 * e^{-\eta t}$ (Nadarajah and Haghighi 2014), with $\eta = 0.714$ and $\eta = 0.781$, for APS and DBLP, and $A_0$ and $Y_0$ are fitting constants. We also perform KS statistic tests, which shows that the goodness-of-fit values are $D_{APS} = 0.25$ and $D_{DBLP} = 0.375$, and *p*-values are $p_{APS} = 0.716$ and $p_{DBLP} = 0.215$, indicating that it can be accepted that the collective attraction of scientific publication will generally decay exponentially. In addition, the exponential form of *r* indicates that papers' influence decay quickly, resulting possibly from fast updating of scientific achievements. Furthermore, the high parameters $\eta > 0$ suggest statistically that the impact of overall publications will not remain constant or constantly increase, but will decay in an exponential form. Moreover, different values of $\eta$ for *APS* and *DBLP* may indicate that decay rates vary by discipline, corresponding to different attractions according to the field.

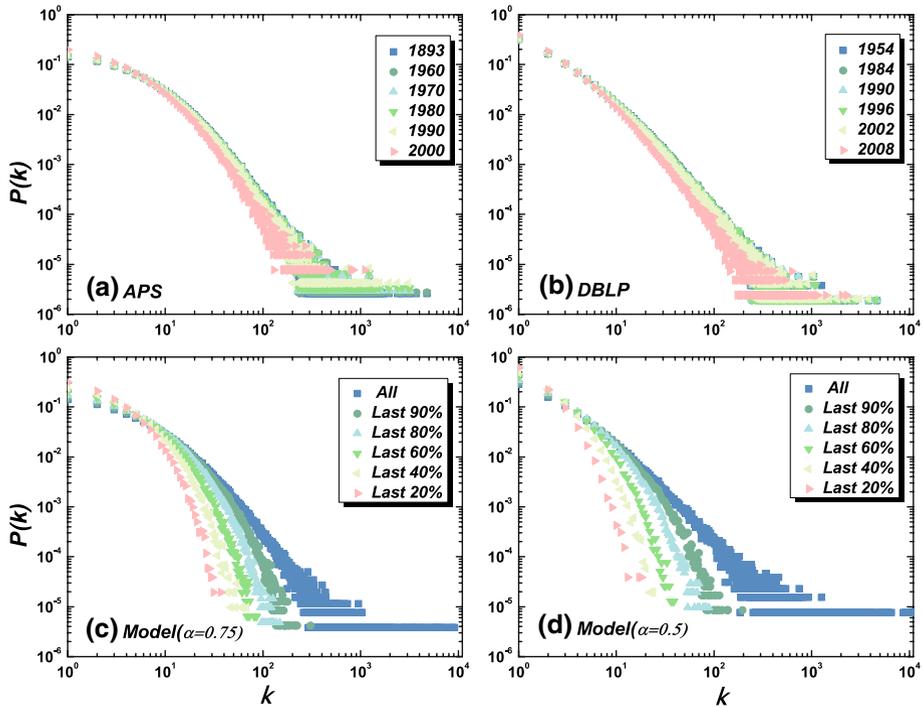

**Fig. 5** (Color online) Comparisons of hyperdegree distributions counting from different times between empirical data and the model. **a** APS data counting from different publication years; **b** DBLP data counting from different publication years; (c-d) Simulation results counting from different *birth* times. Simulation results obtained with **c** $\alpha = 0.75$, and **d** $\alpha = 0.5$

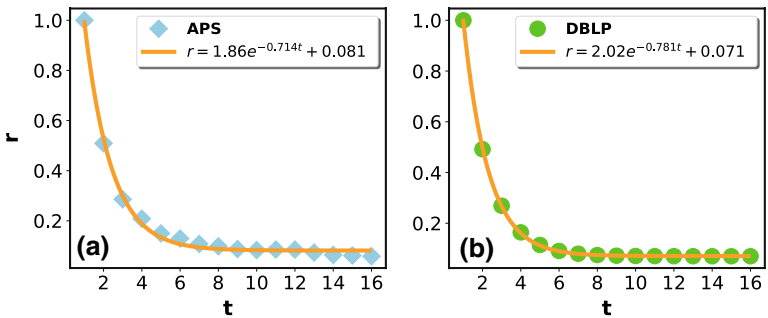

**Fig. 6** (Color online) The decay factor $r$ as a function of time for (a) APS and (b) DBLP. Blue and green dots represent results of the corresponding datasets. Solid orange lines show fitting results in an exponential decay form, $r = Y_0 + A_0 * e^{-\eta t}$, with (a) $\eta = 0.714$ and (b) $\eta = 0.781$. The time interval is set as two years for both datasets

## Conclusions and discussion

We have studied aging effect on the evolution of hypergraph-based citation networks. Empirical analyses from two widely used datasets, *APS* and *DBLP* publications, show that the hyperdegree distribution is significantly affected by the aging factor. We then proposed an evolving model based on a hypergraph structure to illustrate the temporal citation dynamics. The results from both analytical and simulation approaches have been validated using the empirical data. In addition, the experimental results showed that the citation distribution will evolve over different spans of time. Further analysis showed that the decay speed of early publications approximates an exponential form, $r = Y_0 + A_0 * e^{-\eta t}$.

Note that, a citation network can be modelled either as a normal graph or hypergraph. These two models have their own advantages and defects in describing citation relationship. On one hand, a hyperedge is usually treated as an arbitrary number of pairwise links in a normal graph. This model can simplify the complex pattern by using the widely used binary structure to give immediate expression of majority properties. Comparatively, as a hyperedge takes into account the publication and all its references as a whole, it may provide a different perspective to illustrate network characteristics. For instance, it can apparently alter the results of "clustering coefficient" (Estrada and Rodrĺguez-Velzquez 2006) and "network diameter" (Zhang and Liu 2010). On the other hand, many citation patterns can be modeled as directed networks in normal graph. Whereas, studies show that the directed hypergraph (Gallo et al. 1993; Volpentesta 2008) have much more theoretical complexity and difficulties than the undirected one, resulting in much limitation in the applications to describe real-world networks. In addition, the hypergraph theory provides another versatile tool to describe the inherently group structured citation patterns, where such higher-order structure may have crucial dynamical effects (de Arruda et al. 2020). This paper only presents a simply way to unify hypergraph theory, the other network generation mechanisms can also be applied to generate hypergraph network. For example, fitness-based attachment (Bell et al. 2017) and conflicting attachment process (Leung and Weitz 2016), which were used in the normal bipartite network generation model, can be applied to generate new hypergraph network models to characterize high-order relationships on complex systems such as the messaging systems (WhatsApp, WeChat and Facebook Messenger, etc.). In addition, in this paper, we only take into account the power-law decay form as the aging effect, while other functions, e.g. exponential decay, may also be considered as the potential aging effect and need further examination. Furthermore, the underlying dynamics to understand the structure and function of hypergraph based citation networks is worth of focusing on in the future work.

The findings of this work may have various applications in the study of the structure and dynamics of scientific networks. (i) Despite all of the observed data coming from the domains of physical and computer science, the results can be extended to other fields due to possible universal citation patterns; (ii) how scholars trace research fields and how new topics emerge from former disciplines still constitute a major challenge for both the social and natural sciences. The study of time-based citation patterns may provide a promising way to understand the subject from both the micro and macro perspectives. Therefore, to achieve an in-depth understanding of citation dynamics warrants further efforts to develop a more comprehensive model.

# Appendix

### Mathematical analysis of hyperdegree distribution

We attempt to obtain the analytical solution of the hyperdegree distribution based on the master equation. In Eq. (1), we denote $\Omega(t) = \sum_j k_j(t)\tau_j^{-\alpha}$, indicating the contribution of all nodes in the system at time $t$. For $\alpha > 0$, we can obtain that $\lim_{\tau \to \infty} \tau^{-\alpha} = 0$, leading to the convergence value of $\Omega(t)$, $\Omega^* = \lim_{t \to \infty} \Omega(t)$, which is a constant. As a consequence, we can obtain the hyperdegree distribution as

$$p(k;i,\tau+1) = \left(1 - M\frac{k\tau^{-\alpha}}{\Omega^*}\right)p(k;i,\tau) + M\frac{(k-1)\tau^{-\alpha}}{\Omega^*}p(k-1;i,\tau), \quad (3)$$

where $p(k;i,\tau)$ is the probability that node $i$ with age $\tau$ has hyperdegree $k$, and $M = \sum_{m=1}^{L} mq_m$ is the expected number of old papers selected as the "old nodes" at each time step. To depict the aging influence of the citation network, we assume that $L$ and $q_m$ are given; hence, $M$ is fixed. The first term in Eq. (3) is the probability of not selecting papers with $k$ citations, and the second term is the probability of selecting papers with $k-1$ citations. Thus, the fraction of nodes with age $\tau$ and hyperdegree $k$ is

$$p(k,\tau) = \sum_{i \in V} p(k;i,\tau)/\tau, \quad (4)$$

where $V$ is the set of nodes at the corresponding time step.

Summing up Eq. (3) over $i$ through all nodes, we can obtain

$$(\tau+1)p(k,\tau+1) = (1 - rk\tau^{-\alpha})\tau p(k,\tau) + r(k-1)\tau^{1-\alpha}p(k-1,\tau), \quad (5)$$

where $r = \frac{M}{\Omega^*}$ is a constant.

Let $p_k(\tau) = p(k,\tau)$ when $\tau \to \infty$. For the stationary state, we can obtain the following differential equation:

$$\tau\frac{dp_k(\tau)}{d\tau} + (1 + rk\tau^{1-\alpha})p_k(\tau) = r(k-1)\tau^{1-\alpha}p_{k-1}(\tau), \quad (6)$$

with the boundary conditions that $p_k(1) = 1$ for $k = 1$, whereas for $k > 1$, $p_k(1) = 0$ and $p_k(0) = 0$.

Following the method in Ref. (Newman 2009), we can obtain the solution of Eq. (6) as follows:

$$p_k(\tau) = \frac{1}{\tau} exp\left((1 - \tau^{(1-\alpha)})\frac{r}{1-\alpha}\right) \left(1 - exp\left((1 - \tau^{(1-\alpha)})\frac{r}{1-\alpha}\right)\right)^{k-1}. \tag{7}$$

Eq. (7) gives the general solution for the probability distribution of one paper's citations with age $\tau$. The overall distribution of citations over the age from 1 to $\tau_0$, denoted as $P_k(\tau_0)$, can be calculated as

$$P_k(\tau_0) = \frac{1}{\tau_0} \int_1^{\tau_0} p_k(\tau) d\tau$$

$$= \frac{1}{\tau_0} \int_1^{\tau_0} \frac{1}{\tau} exp\left((1 - \tau^{(1-\alpha)})\frac{r}{1-\alpha}\right) \tag{8}$$

$$\left(1 - exp\left((1 - \tau^{(1-\alpha)})\frac{r}{1-\alpha}\right)\right)^{k-1} d\tau$$

Assuming $u = exp\left((1 - \tau^{1-\alpha})\frac{r}{1-\alpha}\right)$, Eq. (8) can be written as

$$P_k(\tau_0) = -\frac{1}{\tau_0}\frac{1}{r} \int_1^{u_0} \left(1 - \frac{1-\alpha}{r} ln u\right)^{-1} (1-u)^{k-1} du. \tag{9}$$

Using the Taylor expansion $1 - \frac{1-\alpha}{r} ln u \approx u^{-\frac{1-\alpha}{r}}$, we can obtain

$$P_k(\tau_0) \approx -\frac{1}{\tau_0}\frac{1}{r} \int_1^{u_0} u^{\frac{1-\alpha}{r}} (1-u)^{k-1} du, \tag{10}$$

where $u_0 = exp\left((1 - \tau_0^{1-\alpha})\frac{r}{1-\alpha}\right)$.

With the substitution $q = 1 - u$, Eq. (10) can be rewritten as

$$P_k(\tau_0) \approx \frac{1}{\tau_0}\frac{1}{r} \int_0^{1-u_0} q^{k-1}(1-q)^{\frac{1-\alpha}{r}} dq. \tag{11}$$

$P_k(\tau_0)$ is the regularized incomplete beta function (Gautschi 1967) and $u_0 = exp\left((1 - \tau_0^{1-\alpha})\frac{r}{1-\alpha}\right)$; hence the final hyperdegree distribution can be approximately written as

$$P_k(\tau_0) \approx A \frac{1}{k}\left(1 - exp\left((1 - \tau_0^{1-\alpha})\frac{r}{1-\alpha}\right)\right)^k. \tag{12}$$

**Acknowledgements** We thank Prof. Pak Ming Hui and Dr. Junming Huang for their invaluable suggestions. This work was partially supported by Zhejiang Provincial Natural Science Foundation of China (Grant Nos. LR18A050001 and LR18A050004), the National Natural Science Foundation of China (Grant Nos. 61663041, 61673151 and 61873080), and the Major Project of The National Social Science Fund of China (Grant No. 19ZDA324).